\documentclass[aps,prl,twocolumn,superscriptaddress]{revtex4-1}

\usepackage{graphicx}
\usepackage{amsmath}
\usepackage[version=3]{mhchem}

\begin{document}

\graphicspath{{Figures/}}

\title{Low frequency spin dynamics in XY quantum spin ice \ce{Yb_2Pt_2O_7}}


\author{S. K. Takahashi}
\affiliation{Department of Physics and Astronomy, McMaster University, Hamilton, Ontario, L8S 4M1, Canada}

\author{A. Arsenault}
\affiliation{Department of Physics and Astronomy, McMaster University, Hamilton, Ontario, L8S 4M1, Canada}

\author{C. Mauws}
\affiliation{Department of Chemistry, University of Winnipeg, Winnipeg, Manitoba, R3B 2E9, Canada}
\affiliation{Department of Chemistry, University of Manitoba, Winnipeg, Manitoba, R3T 2N2 Canada}

\author{A. M. Hallas}
\affiliation{Department of Physics and Astronomy, McMaster University, Hamilton, Ontario, L8S 4M1, Canada}
\affiliation{Department of Physics and Astronomy, Rice University, 6100 Main MS-61 Houston, Texas, 77005-1827}

\author{C. Sarkis}
\affiliation{Department of Physics, Colorado State University, Fort Collins, CO 80523}

\author{K. A. Ross}
\affiliation{Department of Physics, Colorado State University, Fort Collins, CO 80523}

\author{C. R. Wiebe}
\affiliation{Department of Physics and Astronomy, McMaster University, Hamilton, Ontario, L8S 4M1, Canada}
\affiliation{Department of Chemistry, University of Winnipeg, Winnipeg, Manitoba, R3B 2E9, Canada}
\affiliation{Department of Chemistry, University of Manitoba, Winnipeg, Manitoba, R3T 2N2 Canada}
\affiliation{Canadian Institute for Advanced Research, 661 University Ave, Toronto, Ontario, M5G 1M1, Canada}

\author{M. Tachibana}
\affiliation{National Institute for Materials Science, 1-1 Namiki, Tsukuba, 305-0044, Japan}

\author{G. M. Luke}
\affiliation{Department of Physics and Astronomy, McMaster University, Hamilton, Ontario, L8S 4M1, Canada}
\affiliation{Canadian Institute for Advanced Research, 661 University Ave, Toronto, Ontario,  M5G 1M1, Canada}

\author{T. Imai}\email{imai@mcmaster.ca}
\affiliation{Department of Physics and Astronomy, McMaster University, Hamilton, Ontario, L8S 4M1, Canada}
\affiliation{Canadian Institute for Advanced Research, 661 University Ave, Toronto, Ontario, M5G 1M1, Canada}


\date{\today}

\begin{abstract}
The XY pyrochlore \ce{Yb_2Ti_2O_7}, with pseudo spin 1/2 at the Yb$^{3+}$ site, has been celebrated as potential host for the quantum spin ice state. The substitution of non-magnetic Ti with Pt gives \ce{Yb_2Pt_2O_7}, a system with remarkably similar magnetic properties.   
The large nuclear gyromagnetic ratio ($\gamma_{n}/2 \pi = 9.15$~MHz/T) of $^{195}$Pt makes \ce{Yb_2Pt_2O_7} an ideal material for NMR investigation of its unconventional magnetic properties.  Based on the $^{195}$Pt nuclear spin-lattice relaxation rate $1/T_1$ and the magnetic specific heat $C_{p}$ measured in a broad range of magnetic field $B_{ext}$, we demonstrate that the field-induced magnon gap linearly decreases with $B_{ext}$ but additional low energy mode of spin excitations emerge below $\sim 0.5$~T.  

\end{abstract}

\pacs{}

\maketitle


\section{ I. Introduction}

\begin{figure}
	\begin{center}
		\includegraphics[width=3in]{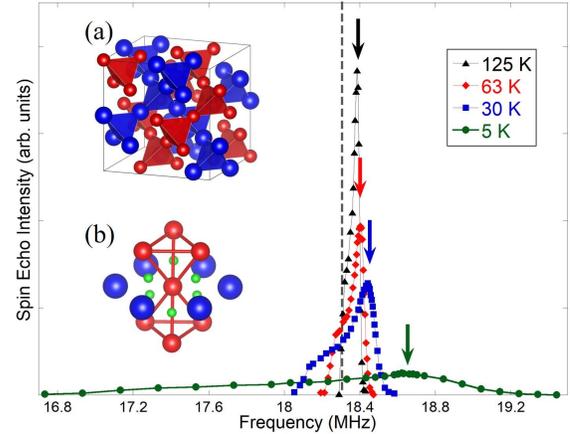}
		\caption{Representative $^{195}$Pt NMR lineshapes measured at 2~T for various temperatures.  Downward arrows mark the main peak for the field orientation of  \(\perp\), whereas shoulders that move to lower frequencies at lower temperature arise from the \(\parallel\) orientation.  Dashed line represents the unshifted frequency.  Inset: (a) The crystal structure of \ce{Yb_2Pt_2O_7} consisting of two networks of interwoven corner-sharing \ce{Yb^{3+}} (blue) and \ce{Pt^{4+}} (red) tetrahedra. Oxygens are not shown for simplicity. (b) The local environment of the platinum sites where the oxygen is in green. When \(B_{ext}\) is in plane with the hexagon of \ce{Yb^{3+}}, we define it as perpendicular ($\perp$).  When \(B_{ext}\) is out of plane, it is defined as parallel ($||$).  }
		\label{fig:Lineshape}
	\end{center}
\end{figure}

Pyrochlore oxides of the form \ce{A_2B_2O_7} have garnered intense interest over the past two decades.  The A and B sites form lattices of corner-sharing tetrahedra, interwoven with each other as shown in Fig. \ref{fig:Lineshape}(a).  Magnetic moments on the A site can be geometrically frustrated, and predicted to create novel states of magnetism such as quantum spin liquids \cite{Moessner1998a, Moessner1998b, Balents2010, Gardner2010}.  Examples of the pyrochlore systems studied to date include classical spin ices (A = Ho, Dy, B = Ti) \cite{Harris1997, Bramwell2001, Ramirez1999}, spin glasses (A = Y, Tb, B = Mo) \cite{Greedan1986, Gaulin1992}, and order by disorder in \ce{Er_2Ti_2O_7} \cite{Bramwell1994, Champion2003, Savary2012, Ross2014}. 

Among the pyrochlore materials of strong current interest is \ce{Yb_2Ti_2O_7} \cite{Gardner2010,Hallas2018}.   The Yb$^{3+}$ site is under the influence of a crystal field, and  has a pseudo spin 1/2 with a larger dipole moment in the XY direction compared to the (local) Ising direction \cite{Hodges2001, Hodges2002,Gaudet2015}.  These frustrated pseudo spin 1/2's at Yb$^{3+}$ sites are a prime candidate for an exotic state of matter that has been sought after for years, {\it the quantum spin ice} \cite{Gingras2014, Savary2012Coulombic} with exotic spin excitations.  In fact, recent experimental studies revealed highly non-trivial spin excitations, such as quasiparticle breakdown effects in low magnetic fields \cite{Bonville2004, Ross2011, Thompson2017, Pan2014, Pan2015, Robert2015, Pecanha2017}.  The nature of the ground state in \ce{Yb_2Ti_2O_7} and its low energy spin excitations are under intense debate. 

NMR measurements are often useful in the investigation of low energy spin excitations in frustrated magnets \cite{Imai2016}, but have played only limited roles in the investigation of pyrochlores \cite{Kaibuchi, NMR}.  In particular, NMR investigation of \ce{Yb_2Ti_2O_7} would require low field measurements and the small values of the nuclear gyromagnetic ratio $\gamma_{n} / 2 \pi \simeq 2.4$~MHz/T of \ce{^{47, 49}Ti} makes such NMR measurements very difficult.  

Very recently, Cai {\it et al}. reported that \ce{Yb_2Pt_2O_7} exhibits very similar properties as \ce{Yb_2Ti_2O_7} \cite{Cai2016}.   \ce{Yb_2Pt_2O_7} orders ferromagnetically at $T_{C} \simeq 0.3$~K as evidenced by a sharp peak of zero field specific heat $C_{p}$, and exhibits a broad hump of $C_{p}$ centered around 2 K.  These observations for \ce{Yb_2Pt_2O_7}  are nearly identical with those reported for \ce{Yb_2Ti_2O_7}.  The unconventional magnetic properties of \ce{Yb_2Ti_2O_7}, such as these specific heat anomalies, have been the subject of intense debate.  We report the first \ce{^{195}Pt} NMR investigation of \ce{Yb_2Pt_2O_7}.  \ce{^{195}Pt} nuclear spins have two major advantages over  \ce{^{47,49}Ti}: the nuclear gyromagnetic ratio $\gamma_{n}/2 \pi = 9.15$~MHz/T is relatively large, and its nuclear spin $I=1/2$ causes no nuclear quadrupole splitting.  We were able to probe the low frequency (NMR frequency from $\sim 3$ MHz to $\sim 85$ MHz) spin dynamics of Yb$^{3+}$ down to $\sim 1.5$~K in a broad field range as low as $B_{ext}\sim 0.35$~T.  We will demonstrate that low energy spin excitations emerge in the low field range below $\sim 0.5$~T. 

\section{II. Experimental Methods}

 We syntehsized the \ce{Yb_2Pt_2O_7} powder crystal under high temperature high pressure (HTHP) conditions at NIMS. A detailed description of the sample can be found in Ref \cite{Hallas2016Relief}.  We conducted \ce{^{195}Pt} NMR measurements at McMaster University using a state-of-the-art spectrometer built around the Apollo NMR console acquired from Tecmag Inc.   All NMR lineshape measurements used the spin echo pulse sequence.  In this sequence, we apply a $\frac{\pi}{2}$ pulse, followed by a waiting period of $\tau$ before applying a $\pi$ pulse and subsequently measuring the spin echo signal. typically, we used the $\frac{\pi}{2}$ pulse width $\sim 2-7\mu$s and $\tau \sim 15 \mu$s.  We used the pulse inversion method for 1/T$_1$ measurements, which is a $\pi$ pulse followed by a spin echo sequence. Since \ce{^{195}Pt} is nuclear spin 1/2, we fit the recovery curve to a single exponential of the form $M(t)=M(\infty)-Ae^{-({\frac{t}{T_1}})}$.   We measured the quasi-adiabatic specific heat, $C_P$, at the University of Winnipeg using a Quantum Design PPMS system.

\section{III. NMR Results}

In the main panel of Fig.\ref{fig:Lineshape}, we show representative $^{195}$Pt NMR lineshape measured in a 2~T for a randomly oriented powder sample. The NMR lineshape shows the typical powder pattern of nuclear spin 1/2 with anisotropic NMR frequency shift with axial symmetry.  Fig.\ref{fig:Lineshape}(b) shows the local environment of the \ce{Pt^{4+}} site.  Each \ce{Pt^{4+}} site is surrounded by a plane of six \ce{Yb^{3+}} ions, each separated from the \ce{Pt^{4+}} site by an oxygen atom. This results in an axially symmetric local environment for each platinum site.  Note that the asymmetry seen by the \ce{Pt^{4+}} site is defined locally and is hence not the same direction as most other platinum sites or \ce{Yb^{3+}} sites.  The peak of the lineshape corresponds to \(B_{ext}\) pointing within the plane (defined as $\perp$ hereafter), whereas the shoulder arises from \(B_{ext}\) pointing out of plane (defined as $||$) i.e. pointing out of the hexagon of \ce{Yb^{3+}} ions as seen in Fig.\ref{fig:Lineshape}(b).  As temperature is decreased, the main peak moves towards higher frequency whereas the shoulder moves to lower frequency.  This is because the hyperfine coupling $A_{hf}$ between \ce{Yb^{3+}} spins and \ce{^{195}Pt} nuclear spins is anisotropic, and $A_{hf}^{\perp} > 0$ while $A_{hf}^{\parallel} < 0$.

\begin{figure}
	\begin{center}
		\includegraphics[width=3.2in]{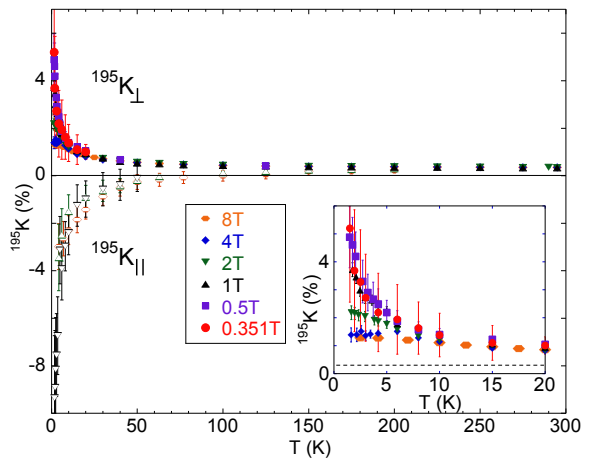}
		\caption{NMR frequency shift at the main peak, \ce{^{195}K}\(_\perp\) (filled symbols) and at the shoulder, \ce{^{195}K}\(_{\parallel}\) (open symbols) as a function of temperature for various fields. Inset shows \ce{^{195}K}\(_\perp\) for temperatures below 20 K.  The dashed line shows the constant offset due to \ce{^{195}K_{chem}} \(\sim 0.34\%\).}
		\label{fig:195Kplot}
	\end{center}
\end{figure}

We deduced the anisotropic \ce{^{195}Pt} NMR frequency shift, \ce{^{195}K}\(_{\perp}\) and \ce{^{195}K}\(_{\parallel}\), from the peak and shoulder of these lineshapes as summarized in Fig. \ref{fig:195Kplot}.  From the results at higher temperatures, we found that the shifts have a constant offset of \ce{^{195}K_{chem}} \(\sim 0.34\%\) due to the chemical shift.  Moreover, we found that the temperature dependent component of \ce{^{195}K}\(_{\perp}\) and \ce{^{195}K}\(_{\parallel}\) shows identical behavior as the powder averaged uniform magnetic susceptibility $\chi$ at $\bf{q=0}$.  Therefore we can express the temperature dependence of \ce{^{195}K} as $^{195}K = A_{hf}\chi/N_{A}\mu_{B} + ^{195}K_{chem}$,  $N_{A}$ is Avogadro's number, and $A_{hf}^{\perp} = 0.73$~kOe/$\mu_{\text{B}}$ and $A_{hf}^{\parallel} = -1.8$~kOe/$\mu_{\text{B}}$.  At low fields and temperatures, the shoulder becomes difficult to decipher from the lineshape, which is why there is no data of $^{195}$K$_{\parallel}$ shown below 1 T.  

\begin{figure}
	\begin{center}
		\includegraphics[width=3.2in]{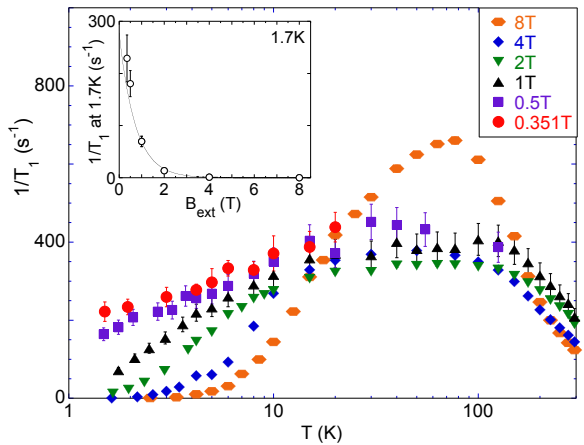}
		\caption{\(1/T_1\) as a function of \(T\) measured at the main peak.  Inset shows the magnetic field dependence of $1/T_{1}$ observed at 1.7~K.  Solid line shows $1/T_{1} \sim exp(\Delta_{T_{1}}/k_{B}T)$, where we extrapolated the observed field dependence of $\Delta_{T_{1}}$ in the high field regime linearly to below 1~T using the fit in Fig. \ref{fig:DeltaPlot}.}
		\label{fig:T1plot}
	\end{center}
\end{figure}

We summarize 1/T$_1$ data in Fig. \ref{fig:T1plot}.  The lower end of our field range is limited to 0.351~T at this time due to the low frequency cut-off of our NMR power amplifiers.    \(1/T_1\) probes the fluctuations of the hyperfine magnetic fields, and may be written as the weighted wave vector \({\bf q}\)-integral of the dynamical structure factor $S({\bf q},\omega_{n})$ of the \ce{Yb^{3+}} spins as $1/T_{1} \propto \Sigma_{{\bf q}} |A({\bf q})|^{2} S({\bf q},\omega_{n})$, where $\omega_{n} = (1+^{195}K)\gamma_{n}B_{ext}$ is the NMR frequency; by ignoring higher order terms, we may represent the hyperfine form factor as $A({\bf q}) \simeq \Sigma_{j=1 \sim 6} A_{NN} \cdot exp(-i{\bf q} \cdot {\bf r}_{j})$, where $A_{NN}$ is the transferred hyperfine interaction from six nearest neighbor Yb$^{3+}$ sites, and ${\bf r}_{j}$ is their position vector with respect to the \ce{Pt^{4+}} sites.  Thus, \(1/T_1\) probes the \({\bf q}\)-integral of what inelastic neutron scattering would observe at the extremely low energy transfer of $\hbar \omega_{n}$($\sim 0.01~\mu$eV).   We note that the single exponential fit of the nuclear spin recovery curve was always good, and hence our \(1/T_1\) results represent the uniform bulk behavior of \ce{Yb^{3+}} spin fluctuations.  We confirmed at fields of 1 T and higher that \(1/T_1\) measured for the \(\parallel\) direction is very similar to that measured at \(\perp\).   

\begin{figure}
	\begin{center}
		\includegraphics[width=3.2in]{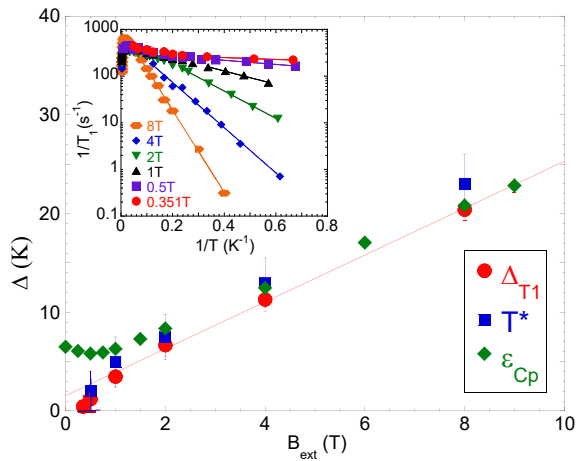}
		\caption{The magnetic field dependence of the gap $\Delta_{T_1}$ extracted from the exponential fit of $1/T_{1}$ as shown in the inset.  Also shown are $T^{*}$ as defined in the inset of Fig. \ref{fig:T1T}, and  the specific heat energy scale $\epsilon_{C_{p}}$. Solid line is a linear fit of $\Delta_{T_1}$ above 1~T, which  overestimate the data at 0.5~T and below.}
		\label{fig:DeltaPlot}
	\end{center}
\end{figure}

As seen in the inset of Fig. \ref{fig:DeltaPlot}, \(1/T_1\) clearly shows thermally activated behavior with decreasing temperature below about 50 K in the field range at least down to $\sim 1$~T.  This indicates a field induced gap $\Delta_{T_{1}}$ in the spin excitation spectrum, and should be attributed to the minimum energy required to excite a magnon in the field-induced spin polarized state.  Earlier inelastic neutron scattering and THz spectroscopy measurements on the sister compound \ce{Yb_2Ti_2O_7} showed the presence of both single and double magnon bands, that are shifted to increasingly larger energies with field \cite{Ross2011, Thompson2017, Pan2014}.    \(1/T_1\) shows a broad maximum between 10 K and 100 K which then decreases at higher temperatures.  This is probably due to the fact that the upper limit of the magnon excitation spectrum is located at $\sim 5$~meV in this field range \cite{Thompson2017}.  

We estimated the field induced gap $\Delta_{T_{1}}$ by fitting the \(1/T_1\) results to $1/T_{1} \sim e^{-\Delta_{T_{1}}/T}$, as summarized in the main panel of Fig. \ref{fig:DeltaPlot}.  Our results indicate that \(\Delta_{T_{1}}\) decreases linearly with $B_{ext}$ down to 1 T, but the gap appears to diminish below 1 T.  From the slope $g\mu_{B}S$ of the linear fit of \(\Delta_{T_{1}}\) above 1 T in Fig. \ref{fig:DeltaPlot}, we estimated $g \simeq 3.4$ for Yb$^{3+}$, in good agreement with $g \simeq 3.55$ estimated by Cai {\it et al}.  based on the magnetic susceptibility \cite{Cai2016}.   Our high field results are qualitatively similar to the case of transverse field Ising chain CoNb$_2$O$_6$, but  $\Delta_{T_{1}}$ decreases linearly to zero in the latter \cite{Kinross}.

\begin{figure}
	\begin{center}
		\includegraphics[width=3.2in]{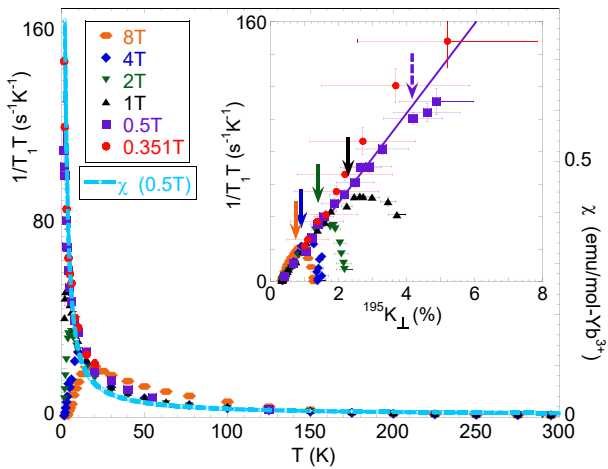}
		\caption{\(1/T_{1}T\) as a function of \(T\). Notice that the downturn due to the field-induced magnon gap at low temperatures is absent below 0.5 T.  We overlay the results of SQUID measurements of the \ce{Yb^{3+}} moment at 0.5 T using a dashed line.   (Inset) \(1/T_1T\) plotted as a function of $^{195}K_{_\perp}$ choosing temperature as the implicit parameter. Solid line represents the best linear fit to 0.5 T data.  For fields of 1 T and higher, linearity breaks down with decreasing temperature due to opening of a gap.  Downward arrows define T*, where \(1/T_1T\) deviates 10\% from highest data point of linearity. The dashed arrow represent T* for 0.5 T, which may be absent.}
		\label{fig:T1T}
	\end{center}
\end{figure}

Also shown in Fig. \ref{fig:T1T} is the temperature dependence of \(1/T_1T\) (\(1/T_1\) divided by \(T\)), which represents the \({\bf q}\)-integral of the imaginary part of the dynamical electron spin susceptibility, Im[\(\chi({\bf q},\omega_{n})\)], at the NMR frequency $\omega_{n}$.  \(1/T_1T\) is peaked at $\sim 18$~ K in 8~T, below which low frequency Yb$^{3+}$ spin fluctuations are suppressed by the presence of a finite gap.  As the field is decreased, the peak shifts to lower temperatures until it disappears at 0.5 T.  \(1/T_1T\) continues to increase rapidly with decreasing temperature in 0.5 T as well as in 0.351 T.  In view of the similarity between the temperature dependence of \(1/T_1T\) and $\chi$ (shown with the right axis), we may attribute the continuing growth of \(1/T_1T\) observed below $\sim 0.5$~T to low frequency Yb$^{3+}$ spin fluctuations enhanced around the ${\bf q}={\bf 0}$ ferromagnetic mode.  

To illustrate this point more clearly, we plot \(1/T_1T\)  vs. $^{195}K_{\perp}$ choosing $T$ as the implicit parameter in the inset of Fig. \ref{fig:T1T}.  We show a linear fit of the data points observed at 0.5~T.  The horizontal intercept corresponds to $K_{chem} \sim 0.34$~\% derived above from the offset in Fig. \ref{fig:195Kplot}.  \(1/T_1T\) increases roughly linearly with \ce{^{195}K} due to the growth of ferromagnetic correlations.    This linearity, however, breaks down at low temperatures in higher fields due to the field induced gap.  We can  define a characteristic temperature, T*, where $1/T_{1}T$ and $^{195}K$ deviate from linearity by 10\%. T* summarized in Fig. \ref{fig:DeltaPlot} exhibits nearly identical behavior as $\Delta_{T_{1}}$, adding confidence that the key features of our observation does not depend on the details of our model analysis.   We also note that the thermally activated behavior of  \(1/T_1\) is barely observable at 0.351~T and 0.5~T in Fig. \ref{fig:T1plot}  and the inset of Fig. \ref{fig:DeltaPlot}.  Moreover, the breakdown of the linearity is hardly observable in Fig. \ref{fig:T1T} at these low fields.  Accordingly, we should consider the results of $\Delta_{T_{1}}$ for these low fields as the upper bound.  

\section{IV. Specific Heat Results}

Thermodynamic measurements provide us with additional insight.  Despite their different ordering behaviors, the Yb$_2$B$_2$O$_7$ (B = Ti, Ge, Sn, Pt) pyrochlores all poses similar magnetic specific heat behaviors \cite{Dun2014,Hallas2016Universal, Hallas2018, Cai2016}. This consists of a sharp lower temperature peak typically below $1$~K indicative of long range order, and a broad peak around 2~K that has been suggested to be the onset of liquid-like correlations \cite{chang2012higgs}.  As noted above, the origin of these specific heat anomalies have been under intense debate over the last decade.  

\begin{figure}
	\begin{center}
		\includegraphics[width=3.2in]{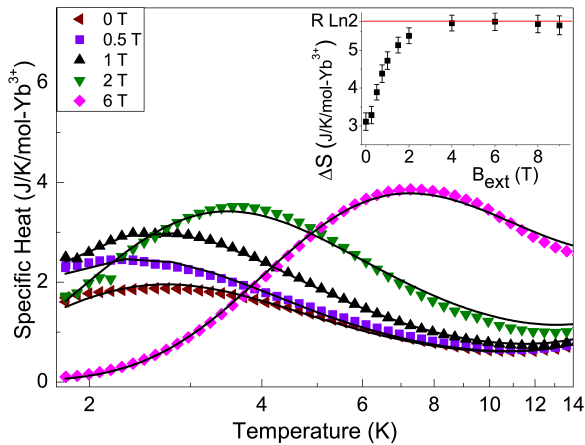}
		\caption{Specific heat $C_{p}$ shown for select magnetic fields.  Solid lines show the two-state system fits used to extract gap energies and entropies.  Notice that the maximum temperature of $C_{p}$ increases from 0.5~T to 0~T.  (Inset) Field dependence of the magnetic entropy release $\Delta S$ associated with the high temperature heat capacity anomaly.
}
		\label{fig:Cp}
	\end{center}
\end{figure}

In Fig. \ref{fig:Cp}, we present the quasi-adiabatic specific heat $C_{p}$ of a pressed polycrystalline 14.1 mg sample.  We removed lattice contributions by fitting a $T^{3}$ component to the zero-field data and fixing it for the finite field results. Our preliminary measurements using a dilution refrigerator at Fort Collins showed the aforementioned sharp lower temperature peak at $T_{C}\simeq 0.3$~K in zero field, in agreement with the earlier report \cite{Cai2016}.  We fit the broad anomaly centered at $\sim 2$~K or above to a two state Schottky model in order to extract field induced gap energies.  A Schottky model fit of the higher field data phenomenologically represents a field induced two state system arising from the field induced magnons with a gap $\epsilon_{C_{p}}$.  Fitting the zero field data to a Schottky model does not necessarily represent that this is a gapped system (in fact neutron scattering measurement have shown otherwise in Yb$_2$Ti$_2$O$_7$ down to a limit of 0.09 meV \cite{gaudet2016gapless}) but provides a qualitatively acceptable fit.  We summarize the deduced energy scale $\epsilon_{C_{p}}$ in Fig. \ref{fig:DeltaPlot}.  The agreement between $\epsilon_{C_{p}}$ and $\Delta_{T_{1}}$ is excellent above 4~T.  Below $\sim 0.5$~T, however, $\epsilon_{C_{p}}$ increases again toward the zero field limit.  

In the inset to Fig. \ref{fig:Cp}, we summarize the field dependence of the entropy loss $\Delta S$ upon cooling through the broad specific heat hump.   In the high field regime down to 4~T, $\Delta S$ is precisely $R \cdot ln2$, as expected for a pseudo spin 1/2 system.  This implies that, under the presence of high magnetic field above 4~T, low frequency Yb$^{3+}$ spin fluctuations completely freeze out in the low temperature range below the hump.  This is consistent with the exponentially decreasing $1/T_1$ due to the field induced gap $\Delta_{T_{1}}$($\simeq \epsilon_{C_{p}}$).  

Below 4~T, $\Delta S$ becomes smaller than $R \cdot ln2$, implying that spins are not completely frozen out at the base temperature.  This is consistent with strong residual Yb$^{3+}$ spin fluctuations reflected on enhanced $1/T_{1}T$ in the low field regime.  In the inset of Fig. \ref{fig:T1plot}, we show that $1/T_1$ at 1.7~K, and hence the residual Yb$^{3+}$ spin fluctuations, grow quickly below $\sim 4$~T.   This could also be related to the persistent spin dynamics seen in muon spin relaxation measurements of this and other frustrated magnetic systems \cite{Dunsiger2011}.  Since $1/T_{1}T$ is roughly proportional to $\chi$, we may use the simple scaling arguments \cite{MMP} to write $1/T_{1}T \propto \Sigma_{{\bf q}} (|A_{||}({\bf q})|^{2}+ |A_{\perp}({\bf q})|^{2})/2 \cdot \chi/ \Gamma$, where we defined the Yb$^{3+}$ spin relaxation rate $1/\Gamma$ to rewrite $Im[\chi({\bf q},\omega_{n})]/\omega_{n} =\pi \chi / \Gamma$.  Using the hyperfine coupling determined from $^{195}K$, we estimate the energy scale as $\Gamma \simeq 0.37$~meV, which is comparable to $\epsilon_{C_{p}}$ as well as the low energy spin excitations observed by neutron scattering in Yb$_2$Ti$_2$O$_7$ \cite{gaudet2016gapless}.   An interesting possibility is that the residual Yb$^{3+}$ spin fluctuations that show up in the energy range below $\epsilon_{C_{p}}$ may be a harbinger of exotic elementary excitations proposed for the quantum spin ice, such as magnetic monopoles \cite{Castelnovo2008}.

\section{V. Summary and Conclusions}

To summarize, we have reported the first NMR investigation of low frequency spin dynamics in \ce{Yb_2Pt_2O_7}.  Owing to the large $\gamma_n$ of $^{195}$Pt, we were able to probe the crucial low field regime of the XY quantum spin ice materials for the first time.  By comparing the NMR results with the specific heat data, we demonstrated that the characteristic energy scale of the field induced excitations $\epsilon_{C_{p}}$ remains finite in the zero field limit, and additional low frequency spin excitations are observed at low temperatures in magnetic fields below $\sim 0.5$~T.  In view of the highly unconventional ordered state observed below $T_{C} \simeq 0.3$~K in a sister-compound \ce{Yb_2Ti_2O_7}, it is tempting to associate the residual spin fluctuations to exotic excitations theoretically proposed for quantum spin ice.  We cannot rule out, however, an alternate possibility that the observed spin fluctuations arise from short range spin correlations that is a precursor of the long range order at $T_{C} \simeq 0.3$~K, especially since short range correlations have been observed well above the transition temperature previously \cite{Bonville2004,Thompson2017}. 

\begin{acknowledgments}
T.\ I.\ and G.\ M.\ L. were supported by NSERC and CIFAR.  C.\ R.\ W.\ was supported by NSERC (DG and CRC Tier II), CIFAR, and CFI.  We created the Fig.\ 1(a,b) of the crystal structure with Vesta \cite{Vesta}.
\end{acknowledgments}



%

\end{document}